Journal of Cybersecurity, 2020, 1–13
doi: 10.1093/cybsec/tyaa006
Research paperResearch paper

# Privacy threats in intimate relationships

Karen Levy [1,2,]* and Bruce Schneier[3,4]

[1]Department of Information Science, Cornell University; [2]Cornell Law School, Ithaca, NY, USA; [3]Berkman Klein Center for Internet and Society, Harvard University; and [4]Belfer Center for Science and International Affairs, Harvard Kennedy School, Cambridge, MA, USA

*Corresponding address: Department of Information Science, Cornell University, Ithaca, NY, USA. Tel: +1 317-682-8287. Email: karen.levy@cornell.edu

Received 11 December 2018; revised 10 March 2020; accepted 8 April 2020## Abstract

This article provides an overview of *intimate threats*: a class of privacy threats that can arise within our families, romantic partnerships, close friendships, and caregiving relationships. Many common assumptions about privacy are upended in the context of these relationships, and many otherwise effective protective measures fail when applied to intimate threats. Those closest to us know the answers to our secret questions, have access to our devices, and can exercise coercive power over us. We survey a range of intimate relationships and describe their common features. Based on these features, we explore implications for both technical privacy design and policy, and offer design recommendations for ameliorating intimate privacy risks.

**Keywords**: intimacy; family; abuse; children; relationships; privacy## Introduction

The information security community tends to focus its attention on a canonical set of attackers: companies tracking our activities online, criminals looking to steal our data, government agencies surveilling us to gather information, and hackers out for the "lulz." But a huge number of threats are much more quotidian, performed by much less powerful and less technically savvy actors with very different motives and resources. These attackers know their victims well, and have much greater access to their information, devices, and lives in general. We call these attacks *intimate threats*, in which one member of an intimate relationship—a spouse, a parent, a child, or a friend, for example—violates the privacy of the other.

Intimate threats have garnered little explicit attention from the security and privacy communities and from system designers. For example, a recent review of 40 academic analyses of smart home security anticipated 29 different threat actors and 100 different types of threats—but the threat model of a domestic abuser was almost entirely absent across the literature [1]. We argue that these threats ought to be treated as a primary concern.

Intimate threats represent the way a huge number of people actually experience insecurity and privacy invasions every day. These threats are so common as to be treated as routine and often overlooked, but they are experienced much more frequently—and often with greater direct impact on victims' lives—than many of the threats that dominate the security discussion. And they disproportionally impact society's most vulnerable and least powerful people, often including women, children, the elderly, and the physically or cognitively impaired. Though these threats are, by their nature, difficult to definitively quantify, the indicators we have suggest the scope and scale of intimate threats. In one survey, 31% of participants admitted to snooping through another person's phone without permission in the past year [2]. A recent Pew survey found that the majority of parents check their teenagers' browsing histories and social media profiles. Forty-eight percent looked through phone records and text messages, and 16% tracked teens' locations via their cell phones; half reported knowing the password to their teenager's email account [3]. An NPR survey of US domestic violence shelters indicated that 85% of shelters had worked with survivors who had been stalked using GPS devices, and that 75% had helped survivors who had been subject to eavesdropping using remote tools [4]. A survey in the UK found that 85% of abuse survivors reported being subject to online abuse as part of a pattern of their abuse more generally [5]. Taken together, figures like these suggest that privacy invasions by intimates are pervasive and deserving of focused study.

In addition to being important on their own, intimate threats can be precursors to more traditional forms of privacy and security threat. Intimate privacy invasions can result in the destruction of valuable or personal data, like financial records or family photographs. They can be the first step in financial fraud. In abusive partner situations, they can be a precursor to physical, emotional, and sexual abuse [6, 7]. And even well-intentioned intimate monitoring can create a slippery slope of acceptability, inuring users to accepting surveillance as a mode of social control in other contexts [8].

©The Author(s) 2020. Published by Oxford University Press. 1
This is an Open Access article distributed under the terms of the Creative Commons Attribution License (http://creativecommons.org/licenses/by/4.0/), which permits unrestricted reuse, distribution, and reproduction in any medium, provided the original work is properly cited.

Finally, a more systematic consideration of intimate threats stands to benefit socio-technical security research as a field. These threats pose difficult technical challenges, made more complex by the social relationships in which they are embedded—which are marked by different degrees of authority and autonomy within relationships. They present a mixture of motivations for privacy invasion, often including beneficent motivations like protection and care. They pose novel and interesting questions about privacy boundaries: what degree of monitoring is socially and normatively acceptable in intimate relationships, and how system designers might best accommodate divergent and dynamic preferences. Directly addressing these issues extends the field and provides designers with an opportunity to better address real-world situations. In this way, our work fits into a broader scheme of research that prioritizes the sociotechnical and behavioral dimensions of security and privacy across different social contexts, and which recognizes the critical importance of interdisciplinary approaches to developing solutions [9–11].

Our goals in this article are twofold. While emerging research has begun to examine privacy threats within particular intimate relationships, we are aware of no work that synthesizes common characteristics or design considerations of these threats from across intimate contexts. Our first goal, then, is to describe intimate threats as a *class* of privacy problems, drawing out the features that characterize the category. Many of these features involve the violation of implicit assumptions that hold more readily in other contexts of privacy threat. A better understanding of these common features is required to more adequately protect against intimate threats.

Our second goal is to articulate a set of design considerations that is cognizant of intimate threats. These are difficult problems, and our intention is not to prescribe an exhaustive "checklist" that will immunize a technological system against all intimate threats. Rather, we aim to supply researchers, designers, and policymakers with a conceptual toolkit for recognizing and taking these threats seriously, as well as a critical assessment of the design trade-offs they entail.

## Monitoring in intimate relationships

An extensive amount of monitoring routinely occurs across many types of intimate relations, from romantic partners, to parent–child relationships, to roommates, to caregivers. Family members, roommates, and close friends often know each other's whereabouts and with whom the other spends time. Long-term partners often share bank accounts and keep track of each other's financial activities. Roommates answer each other's phone calls—regularly on a shared home landline, and sometimes on each other's cell phones. People living in the same household may share computers, phones, and other connected devices. Intimates might share social media and email accounts [12]—and even if they have separate accounts, they may know one another's passwords [13–15]. Depending on how their devices and accounts are configured, they may have access (intentionally or not) to each other's files, browsing history, and more. Smart home devices are shared by necessity, and give family members access to a great deal of information about each other's whereabouts and activities.

People may willingly share access to accounts and devices for a number of benign and useful social, cultural, and economic reasons [12, 16]. They may do so as a practical component of household management and communication [16], or because it is cost-effective to pool resources within the family. They may do so to establish and demonstrate intimacy [17] or trust [18, 19] in a partner, or as a condition of access. Personal preferences and cultural expectations further complicate matters.[1] Some partners may desire not only to monitor an intimate partner, but also to *be* monitored, for convenience (e.g., "I like my partner to know when I'm on my way home so we can make evening plans") [21], for safety (e.g., to inform trusted contacts of one's location to provide a "virtual escort" while walking alone) [22], or for other reasons. In other contexts, there may be social or cultural assumptions of family access and sharing, often along gendered lines [23, 24]. (In fact, some industry groups have gone so far as to say that because devices are often shared within households and families, device identifiers should not be considered "personally identifying" under privacy laws [25].)

Much of this access is not necessarily nefarious, intentional, or even unwelcome. In many cases, it simply reflects how people choose to organize their households and relationships, and the role of digital technologies within them. But intimacy also presents distinct informational vulnerabilities. Those who sit in intimate relation to us hold unique resources that can be brought to bear to gain access to our data or devices. Intimates may marshal those resources for a variety of purposes, up to and including abuse. And even in non-abusive situations, members of close relationships may find it almost impossible to protect their own privacy interests against one another, thanks in large part to assumptions built into common technical infrastructures.

Intimate monitoring brings unique ethical complications to the fore [26, 27]. In most privacy contexts, there's little question about the impropriety of unauthorized access. But in intimate settings, some unauthorized access would strike many as warranted—or even required—as a component of a duty of care [28]. Family members have a moral, economic, and often legal responsibility to take care of one another and ensure each other's safety, security, and well-being; they often leverage data-gathering technologies in doing so. Indeed, there are many cases in which it seems both normatively and practically unfathomable that intimates *not* be privy to one another's data. For instance, medical and educational data from minor children must be made available to their parents—who bear responsibility for children's care in both respects—and applicable laws specifically provide that parents should have such access in many cases.[2]

The fact that intimate information-sharing is widespread and often accepted should not lead us to be unreflective about very real privacy threats within intimate relationships. Rather, it makes it all the more important to consider how intimate privacy threats occur, when they are unwelcome, and how to reason about them conceptually. The line between *watching* and *watching over* is a blurry one. Even in close, loving, and generally functional relationships, privacy invasions can at times be cause for conflict and anxiety [31]. There are no bright-line rules for determining when duties of care and protection override privacy interests in intimate relationships, nor for whether intimate monitoring crosses a line of appropriateness. Intentions can be intertwined in the same monitoring relationship ("I want to see when my wife is on her way home so I can start dinner, but I also want to make sure she isn't going near X's house") and can change over time and circumstance. Variable preferences and norms about data sharing in intimate relationships present special challenges to designing privacy into these systems; they are not a reason to ignore these threats.

---

1　For an example of such monitoring that many would find abhorrent, see this about women in Saudi Arabia [20].

2　For example, in the USA, the Family Educational Rights and Privacy Act (FERPA) gives parents access to their children's educational records up until age 18; even after age 18, schools may *choose* to disclose certain records to parents in some cases (e.g., in cases of an emergency, or if the child is claimed as a dependent) [29]. Indeed, the tension between parental notification and a child's privacy can be a difficult one for institutions to navigate, as when colleges do not notify parents of children's psychological difficulties [30].

Victims of intimate privacy threat typically lack legal recourse. Judges and legislators are generally loath to intervene too strongly in what is often considered the "sacred" space of the intimate sphere, tending to protect the privacy *of* families vis-à-vis the state rather than privacy *within* the family [32–34]. This is even more true in patriarchal societies that grant men in the family stronger rights and freedoms (e.g., Turkey [35]); in Saudi Arabia, for example, where all women are required to have a male guardian, a government-run website permits men to grant or deny women under their guardianship the right to travel, and to set up notifications so that they receive a text message should a woman in their family try to get on an airplane [20]. In the USA, some legal protections do exist against the most egregious abuses—protections against nonconsensual pornography [36, 37], no-contact orders following episodes of intimate partner violence, and so on—but the law has generally had trouble keeping up with the challenges of digitally mediated intimate abuse [38–40]. And in the contexts of intimate privacy violations that do *not* rise to the level of abuse (like child tracking), the law offers virtually no remedy, often due to the assumption that a caretaker's preferences are aligned with the interests of the person being monitored [41, 42].

We conceive of intimate privacy threats broadly in this article. Though we use the words "attack," "attacker," and "victim" to characterize aspects of these threats, these may sometimes seem to describe normal—even accidental—interactions between people with no specific malicious motivations toward one another. Our use of these words aligns our inquiry with the dominant discourse in security discussions and threat modeling, in which the terms merely indicate who is attacking and defending a system and do not have any moral or pejorative connotations [43, 44]. Many of the privacy invasions we discuss in this article are quite casual; attackers need not necessarily act with bad intent, nor plan to use the information gleaned for abusive or illegal purposes. It may be helpful to conceptualize some intimate threats as *involuntary disclosure* of a victim's information to an attacker, at times even without the attacker having a specific intent to obtain that information.

## Types of intimate attackers and victims

All intimate relationships—and the privacy practices and expectations within them—are different. We characterize here some of the most prevalent relationships that may give rise to privacy threats and summarize some of the existing research that examines each context. The scope of intimate relationships with which we concern ourselves follows Hasday's definition of intimates as including "dates, sexual and/or romantic partners, and family members such as spouses, parents, and children" [34, p. 6]; we consider in-family caregivers (e.g., nannies) and friends/roommates, as well. (Hasday argues that no "ideal and unassailable definition of intimacy exists"; like her, we adopt a "working definition" based on common understanding of the term, without firmly fixed boundaries.)

### Intimate partners

Privacy threats commonly emerge in romantic partnerships. Significant others may invade one another's privacy for a variety of reasons, ranging from casual to abusive, over the course of a romantic relationship.

Perhaps the most alarming example of privacy invasion in a romantic relationship is in the case of intimate partner violence and abuse. Nearly one in three women and one in six men will experience abuse at some point over the course of their lives [45]. In an increasing proportion of these cases, abusers use digital tools to perpetuate abuse and control over the victim: tracking a victim's location, monitoring their communications, harassing and threatening them, and otherwise surveilling or restricting their activities [6, 7, 46, 47]. These behaviors commonly begin in early dating relationships, and often accompany or prefigure other forms of abuse [47].

Abusers have used a variety of digital tools to spy on or exert control over their victims, most of which require minimal technical sophistication. Some abusers use off-the-shelf spyware apps, which are commonly available online and on app stores; these applications run in the background on victims' mobile phones and computers, reporting their activities back to an abuser surreptitiously [48, 49].

Abusers also make use of a number of commonplace digital tools for abusive purposes. Smart home and IoT technologies such as remote web-enabled cameras, home appliances, thermostats, speakers, and other home sensors can be used by abusers to stalk, harass, and monitor victims and their activities [50, 51]. One company markets a mattress that detects and reports "suspicious movements in the bed" [52].

Even more commonly, abusers rely on the ease of access facilitated by knowledge of the victim's passwords (sometimes shared under threat, other times voluntarily as a sign of trust [19, 53]), answers to security questions, and other authentication mechanisms. Sometimes authentication can by bypassed altogether—for example, the abuser may be able to keep track of the victim's communications because they share a family plan for cellular service, or via browser history on a shared computer [46, 54]. Child-tracking apps and employee-tracking apps are easily repurposed for surreptitiously monitoring intimate partners, and there are some indications that their developers are aware of and condone such use [48]. Social media also presents an easy route toward tracking, as many platforms offer up information like a user's location while posting or an indicator of whether the user is actively on the site. Many shared services record usage history, which can be used to monitor a partner.

Privacy invasions in the context of intimate partner abuse are especially egregious and provide a noncontroversial (and important) rallying point for taking intimate threats seriously. But privacy invasions between romantic partners aren't restricted to these extreme cases, and we ought to take intimate privacy seriously even in the *absence* of abusive circumstances. For instance, many partners collect data about one another routinely and harmlessly in the context of courtship (e.g., "Facebook stalking" a prospective date), sexual relationships, relationship management, or as a way to "gamify" aspects of romantic love [17, 32, 55, 56]. Many fertility- and pregnancy-focused apps grant partners a degree of surveillance over one another (most commonly, male partners over female partners) [57]. For example, one service prompts a man to offer his pregnant partner a glass of water if his version of the app suggests she may be dehydrated [32]. Several period trackers issue "alerts" to men when their partners are menstruating; one even provides the capability to track several women's periods at once, tracking each with a separate password "so when you punch it in, it only looks like you're tracking her" [32].

In other cases, partners may monitor each other when they doubt the responsibility of each other's behavior; for example, parents who travel commonly report "checking in" on their partner via web-enabled baby monitor to see if the baby has been put to bed properly. Soberlink, a facial-recognition-augmented breathalyzer, is sometimes ordered by courts as a condition of visitation when one parent has a history of alcoholism. If the alcoholic parent (the "monitored client") fails to breathe clean, a text is sent to the other partner (the "concerned party") as an indication that it is not safe for the children to visit that day [58].

Additionally, partners are sometimes "caught" being unfaithful via monitoring unbeknownst to them. For example, the governor of Alabama's furtive texts to his paramour were being synced to his wife's iPad [59]. In another story, an Internet-connected smart scale sent the weight measurement of someone's illicit lover to his partner's phone [60]. In these cases, there's a tendency to view the unfaithful partner as a villain who had it coming, rather than as a person whose privacy preferences were disrespected by poor design [12]. But from a privacy-protective perspective, we ought to be agnostic as to the nature of the behavior or content detected, and be fundamentally concerned with how technology may facilitate involuntary information-sharing.

### Parents and minor children

Parents routinely monitor their children in the course of caring for them, from infancy (and even beforehand, *in utero*) through adolescence [61]. Some degree of parental monitoring is essential to ensure children's safety and well-being. Indeed, as parents' lives become busier, parents are often lambasted or punished for giving children significant autonomy—a burden disproportionately felt by women of color and at the lower end of the socioeconomic scale. Mothers have been charged with child abuse and endangerment for letting their children wait in the car or play in a park unsupervised while they run errands or attend job interviews [62]. This risk of being perceived as a neglectful parent, combined with a lack of social or governmental infrastructure for providing childcare resources, provides an incentive for parents to digitally track their children. Fear-based marketing exacerbates this impetus by cultivating a sense of generalized anxiety in parents—one that can most readily be ameliorated through monitoring [28].

Monitoring often continues well into the teenage years, as different risks become salient to parents. Many parents know the passwords of their children's accounts and regularly check on their online activities, perhaps as a condition of use [63]. Parents have been held responsible for their children's illegal file downloads or sexting behavior, creating legal obligations that result from *failing to* supervise teens' online activity [64, 65]. As discussed earlier, a Pew survey found that most parents engaged in some form of monitoring of teens' browsing histories and social media profiles, and half had their teens' email passwords [3]. A separate study found that parents with home-entryway surveillance systems routinely monitor the comings and goings of their teenage children [66]. Parental monitoring software is commonly marketed to aid parents in many of these activities [28, 48]. On the more extreme end of the spectrum, parents may purchase tamperproof ankle bracelets and GPS monitoring services for "high-risk" teens [67].

The balance between essential caretaking and privacy invasion can be unclear [68, 69]. On one hand, parents have a duty to supervise their children, and implicit authority to place limits on their activities and communications. Parental control apps like Google's Family Link allow parents to view children's online activity and device location, under the advertised purpose of letting parents "set digital ground rules to help guide" their children online [70]. On the other hand, some have raised concerns that the normalization of parental surveillance quashes developmentally important childhood freedoms and trust-building—particularly as children get older—as well as children's freedom of expression and access to information [71–73]. Child monitoring apps like Bark, for example, alert parents when its algorithms detect profanity, sexting, or indicators of depression in a child's social media or text exchanges [74]. Toys like Hello Barbie record children's conversations with the doll and, unbeknownst to them, email the audio files to their parents [75]. A recent Google patent proposes that its smart home system can "infer mischief" if its audio and motion sensors detect that children are occupying a room—but are *too* quiet [76]. All have prompted scrutiny from privacy researchers.

Parents may also violate their children's privacy for reasons wholly unrelated to caretaking. Parents may fraudulently use a child's identity for purposes of opening lines of credit and other accounts. Though the prevalence of such fraud is difficult to establish empirically, research suggests that when a child's identity is stolen, their parents are the most likely perpetrators [77].

In other contexts, the tables may be turned: young children may be privacy threats *to* their parents. Children are often the savviest technology consumers in their own families, and often act as "sysadmins" within them; in the course of this role, they may incidentally or deliberately gain access to detailed digital information about their parents [78]. And children may have motivations to use this information for personal gain—stealing money from parents' bank accounts, using parents' passwords to gain access to proscribed media, using their credit cards, and the like. Notably, some authentication mechanisms may be less effective for one's children for reasons having to do with biological similarity. The chance of a random person unlocking someone else's Apple's Face ID—used for authentication on the iPhone X—is only one in one million, according to Apple's whitepaper on the topic—but "[t]he probability of a false match is different for twins and siblings that look like you as well as among children under the age of 13, because their distinct facial features may not have fully developed" [79]. Indeed, cases of children unlocking their parents' iPhones with the children's own faces have been reported in the media [80].

### Adult children and elderly parents

As the world's population ages, a growing number of families find themselves charged with caring for elderly relatives [42]. The corresponding demand for care—along with meager state resource allocations to support such care—leave many families dependent on remote monitoring technologies to make these burdens tractable.

Some families use video monitoring equipment, colloquially known as "granny cams," to keep tabs on the safety and well-being of elderly relatives [42, 81]. In nursing homes and assisted living facilities, families often deploy web-enabled cameras in residents' rooms. The use of these cameras is often motivated by concern about the resident being abused or neglected at the hands of staff or another resident [42]. Roughly 10% of elderly adults (across all care settings) are estimated to be victims of physical, sexual, or psychological abuse, neglect, or financial exploitation [82]. And nursing home residents are considered to be among the most vulnerable: approximately half of nursing home residents suffer from Alzheimer's disease or related dementias [83], and abuse is believed to be significantly underreported among populations afflicted with these conditions [84].

But familial monitoring of elderly relatives presents its own set of privacy threats. The same cognitive impairments that make nursing home residents susceptible to abuse may also make them unable to give meaningful consent to being monitored by a relative. When this is the case, the capacity for consent typically defaults to the resident's "representative," who is most commonly the family member who is instigating monitoring in the first place [42]. A huge variety of intimate activities—including bathing, dressing, medical care, sexual activity, and personal conversations—takes place in residents' rooms. Since 2001, seven states have implemented statutes and regulations governing families' use of cameras there—but the majority of such legislation does not account for inconsistent privacy preferences between the resident and the family representative (nor do they account for potential abuse situations within the familial relationship). Instead, they tend to treat the family member's decisions as a precise extension of the interests of the elderly resident [42].

Alternatively, families may monitor an elderly relative to support "aging in place"—that is, as a condition of permitting the relative to remain in a private home, often alone, rather than moving them to a facility where they would have better access to medical services but might lose desired independence [85]. Cameras are also often used in these contexts, as well as a variety of other technologies that give a family member oversight over the activities of the elderly relative. These commonly include monitoring of health outcomes and behaviors, like adherence to prescriptions (like "smart" pills and pill bottles that notify someone if a family member fails to take medicines on time [86, 87]), safety and mobility issues (like Lifeline systems that detect falls [88]), and a variety of smartphone apps, GPS trackers, and in-home sensor systems that track things like temperature, doors opening and closing, and the presence of visitors [89]. The common denominator among such technologies is a rhetoric of enablement: but for the peace of mind that they ensure, the elderly relative would no longer be able to live independently [90]. As is the case in other intimate relationships, family members' monitoring of elderly relatives is very often motivated by care and a desire to protect. Yet, research suggests that the privacy preferences of monitored relatives often diverge. In one study, adult children of elderly mothers had consistently more favorable views of sensor, camera, and location tracking technologies than their mothers did—but the adult children typically thought they could persuade their mothers to give consent to being monitored [91].

### Other caregivers and their charges/patients/dependents

Similar intimate threats arise in the context of paid care work. An increasing amount of intimate care is outsourced to nannies, babysitters, and workers who care for the elderly and infirm. The presence of these workers as intermediaries in care relations introduces further opportunities and incentives for intimate monitoring, as well as additional complexities related to the employment relationship.

These workers may themselves monitor their charges, using the same sorts of tools, and based on the same sorts of motivations, as described above. They may also be the targets of monitoring by their employer (or by a government agency that subsidizes the care)—to ensure that they do their work to a satisfactory level, to allay concerns that they may steal from the household, and to ensure the safety and health of their charges [92]. This monitoring commonly occurs via nanny cams and distributed surveillance platforms like Nannysightings.com, through which parents can report to one another on caregivers' behaviors [28]. Extensive monitoring can also occur in the context of hiring and screening caregivers: the service Predictim, for instance, analyzed prospective babysitters' social media histories in order to predict their propensity for drug abuse and bullying (before Facebook and Twitter curtailed their access to do so) [93].

Besides being attackers or victims of attacks themselves, paid caretakers can also be used as a justification for more monitoring of the dependent by the person who contracts for their care. For example, the threat of abuse at the hands of nursing home workers is used as a justification for putting elderly residents on cameras monitored by family members (despite the fact that most elder abuse is perpetrated by family members, not care workers), potentially resulting in invasions of the elderly resident's privacy by their family [42, 81]. Similarly, some day-care centers offer web cameras for parents to monitor the type and quality of care their children receive (see, for example, [94]).

### Friends

Of course, privacy threats can also arise within friendships. Friends often share intimate details of their lives with each other; in fact, the willingness to reveal private information to one another can be understood as an indicator of trust and closeness in the relation [8]. Friends may be roommates and share common physical space. But as with other sorts of intimate relations, friends can be controlling and retaliatory, and friendships can sour. As such, they can share many of the same characteristics as other intimate relationships.

This class of risk can be further exacerbated by the inexperience and naïveté of youth, and by the transitory nature of friendships and partnerships among teens and tweens [53]. Young people manage, define, and maintain their relationships with one another by differentiating the access they allow to some friends versus others (e.g., allowing some friends—but not others—to know one's location on a "Find My Friends" app) [8].

## Common features of intimate threats

Having reviewed various relations in which intimate threats can reside, we turn now to drawing out features that frequently characterize intimate threats across these relational contexts, as described in existing research—many of which set them apart from traditional privacy contexts. Clearly, individual situations will vary; these features will be present and more or less salient in different relationships. We enumerate four such features here, and in the following section describe their implications for policy and design.

### Feature 1: Attackers may have multiple motivations—including beneficent ones—often tied to emotion

Attacker motivations in intimate settings are often very different than in other privacy contexts. Although there are certainly instances of intimates stealing money and other things of value from each other, in general, intimate attacks are more likely to be motivated by an attacker seeking knowledge of, and possibly control over, another's behaviors [95]. Sometimes these motivations are premised on positive inclinations like love, caretaking, and perceived protection from internal and external dangers. There may be a strongly held (and legally supported) sense of duty to "look after" intimate relations, and privacy invasions may be justified as being "for their own good"—particularly when one party is much more vulnerable, like a child, elderly adult, or a family member with reduced physical or cognitive capacity.

In other cases, the motivation may be control for control's sake, jealousy, or fear. In abusive situations, the motivations may be a desire to cause emotional or even physical harm, retaliation for a perceived wrong, or preventing a victim from seeking help or extrication from the situation [6]. On both ends of the spectrum, emotion plays a strong role in motivating behavior, and advertising often plays on those emotions to market monitoring tools [28].

These emotional motivations mean that normal considerations about whether an attack is "worth it" can fail in the context of intimate relations. Because an attacker may be motivated by a range of factors—ranging from deep love and care to obsession, jealousy, or desire for control, and with a good deal of variation in individual, cultural, and relational preferences—dispassionate, rational cost-benefit analysis of threats and resources is unlikely to be easily applied to intimate threats. One of us (Bruce) remembers that as a child he once brute-forced a combination padlock in his house. A four-digit lock's 10,000 possible combinations might be enough to keep out a burglar, but fail against a child with unlimited access and nothing better to do that day.

## Feature 2: Copresence facilitates device and account access

In many privacy contexts, attackers and victims are assumed, at least implicitly, to occupy physically separate spaces. Physical separation helps to ensure that authentication mechanisms and access credentials create security. This assumption rarely holds true in intimate relationships. We borrow here from Goffman's use of the term *copresence* to describe situations in which two actors share physical space, facilitating "rich[] information flow" between them such that people "are close enough to be perceived in whatever they are doing" [96, p. 17]. In intimate relationships, people very commonly share physical space—they live together in households, spend time together in public and private settings, and otherwise have high degrees of physical access that facilitates information transmission about each other. Copresence has a number of implications for intimate threats, as we describe here.

Shared physical spaces and proximity among threat, victim, and devices create different vulnerabilities than those threats premised solely on remote digital access [46]. Copresence allows attackers to access a victim's devices physically, facilitating information visibility (including "over the shoulder" threats such as reading the victim's screen, watching them enter their passwords, and so on [19, 54, 97]), as well as easier installation of spyware [48]. Many smartphone apps default to presenting messages and communications on the phone's locked screen, a potential vulnerability if a user's intimate also has access to the physical device. Other information may be transmitted through jointly used resources in a shared space: a family might have a single shared computer, or a common backup system for all the household's computers.

Copresence can also reduce the effectiveness of security measures like two-factor authentication. The most common second factor is a smartphone, to which intimate attackers often have at least intermittent access. This can enable them to read any one-time access codes displayed on the locked screen. Copresence can even defeat biometric authentication. In one published incident, a woman unlocked her husband's smartphone by placing his sleeping hand on the fingerprint reader [98].

Further, copresence compounds the forms of attack to which a victim is vulnerable. Unlike a physically distant privacy threat whose access to the victim is entirely digital,[3] an intimate attacker may expose a victim to other forms of attack, like physical, sexual, emotional, and financial abuse. In some cases, to avoid escalation via other abuse vectors, victims' advocates may advise a victim *not* to cut off the abuser's digital access, because doing so can lead to escalation of abuse in other forms [6]. Counterintuitively, then, it may be in the victim's best interest *not* to immediately ameliorate digital threats, or even to indicate their awareness of them.

Finally, because many people are involved in family relationships, attackers may leverage *other* co-present family members in the service of monitoring another. For example, some survivors of intimate partner abuse report that even if they maintain digital security on their own devices, they can be indirectly monitored via devices controlled by a shared child [46].

## Feature 3: Intimate relationships have inherent, dynamic power differentials, backed by explicit or implicit authority

Privacy invasion often accompanies and extends existing vectors of relational power [6]. In many cases, the monitored party has relatively less power in the relation by virtue of age, various forms of dependency (legal, financial, and so on), social norms (men having authority over women in some cultures), or reduced capacity (children, victims of intimate partner violence, elderly adults with dementia, and so on). Intimate threats are very likely the threats most frequently experienced by women, children, and those with disabilities. Power dynamics are also likely to change over time—as the nature of a romantic relationship changes, as children age, as an adult's cognitive abilities decline and he becomes more dependent on caregivers, and so on.

In many cases of intimate threat, the attacker has decision-making authority over the victim: granted either explicitly by law, or implicitly by the design of the system. Examples of explicit authority are parental rights and responsibilities to access a child's data or to vicariously consent to monitoring on that child's behalf [99], or a power of attorney for someone with diminished capacity. This authority may undermine consent-based models of privacy protection: the attacker both has authority to consent on behalf of the victim and *is themselves* a threat to the victim's privacy, creating a circular (and nonprotective) situation [42]. And some legal frameworks explicitly permit or require data sharing between intimates, like the provision of student data to parents under FERPA, court-ordered alcohol monitoring for parental visitation, or state statutes that permit families to record their loved ones in nursing homes.

An attacker's authority may also be implicit, based on ownership or expertise. For example, the person who pays for a phone family plan may have the capability of accessing data for all users. Decisions about installation and use of smart home monitoring systems are often driven by the individual in the house with the most expertise and control over the household; Geeng and Roesner [100] found that these decision-makers often didn't consult other members of the household about these decisions because "they did not consider them equal decision-makers in the home." Power differentials also imply that coercion can be an important enabler of surveillance in intimate relationships. Intimate attackers can coerce or threaten their victims to keep their smartphones unlocked, divulge the passwords to their social media accounts, or enable location tracking [101].

---

3   But see some complications of this in contexts like swatting.

### Feature 4: Attackers may bring deep knowledge resources to bear in order to exploit relational vulnerabilities

Privacy infringements in intimate relationships tend to be technically simple. They can involve no more than using readily available device and account interfaces, and attackers need not have great technical skill to execute attacks. But what these attacks lack in technical sophistication, they make up in relational complexity. Simply because of their extensive knowledge of the victim, intimate attackers have deep relational resources that they can leverage in several ways.

Attackers may use intimate knowledge of the victim to gain access to accounts [46]. Much of this information is shared willingly during a relationship and may be shared without consent afterwards. Intimate social knowledge negates certain forms of authentication, which often rely on knowledge of a person's history and social life, under the assumption that attackers would not have access to such information. Some banks authenticate users by asking them for prior addresses; security questions often seek information like a mother's maiden name, a favorite pet or teacher, or a birthday. These types of information, of course, are commonly shared with one's intimates. One of Facebook's backup authentication systems involves showing the person photographs of people and requiring them to accurately identify the ones they know [102]. This is something an intimate partner or family member can do as well. In one recently publicized incident, an Australian woman's ex-boyfriend stalked her with the assistance of an app integrated with her vehicle, which reported her location to him; because he'd helped her purchase the vehicle, he had access to the car's registration information [103].

What's more, thick relational ties complicate amelioration of privacy threats, and create leverage for the attacker. A distant hacker likely has no knowledge of a person's immigration status, health conditions, or personal "dirt" that can be exposed to others online; an intimate associate has access to all of these [6]. Partners may also have access to intimate photos of each other, enabling revenge porn. Control over a spouse's finances, a child's curfew, or an elderly relative's ability to live at home further gives the intimate attacker control; all of these may be conditioned on intimate monitoring, further complicating consent and amelioration.

## Implications for policy and design

While many of the threats we have described here are technically unsophisticated, we should not misread this as an indication that they are easy to solve. The social complexity and heterogeneity underlying intimate threats make them very challenging to address technically—which is, perhaps, why they are often ignored by engineers and designers. (Other researchers have pointed out the very low proportion of cybersecurity professionals who are women and minorities, and have suggested that this lack of representation may also lead to underemphasis on threats predominantly experienced by those groups [104].). Intimate privacy invasions are often diffuse and covert, unlike the high-profile data breaches regularly reported in the news and may therefore also garner less attention and concern in system design.

Some aspects of this problem must be mitigated by law and policy. A recent Citizen Lab report on stalkerware concluded with a list of detailed policy recommendations to regulate that industry [49]. Further, we need increased penalties for abuse cases that include digital tracking. Eva Galperin of the Electronic Frontier Foundation has called on US law enforcement to prosecute stalkerware companies on hacking charges [105]. Legal scholar Danielle Citron has also articulated a policy agenda to increase civil and criminal penalties against these companies, and to increase digital forensic training for state and local agencies [38]. Some laws have attempted to criminalize the usage of more general IoT devices for surveillance purposes.

The degree to which system designers should be held morally responsible—or legally liable—for every misuse of the technologies they develop is a policy question without easy answers, particularly for general purpose technologies put to unintended uses, and we do not attempt to address it here. However, by taking intimate threat models seriously from the outset, system designers can take some steps to proactively mitigate the risks of intimate partner threats. Importantly, many forms of design may have important roles to play in this mitigation, from visual aspects of a user interface to core system functionalities, and including both the design of physical "things" and of information flows and processes [106].

All engineering involves trade-offs, involving both security and functionality. The same capability that allows a parent to monitor where their child goes online can also allow a spouse to monitor their partner. And some attacks simply can't be detected by technology: a remote website, for example, will very likely not be able to tell when someone is authenticating under the duress of threatened physical violence from an abuser. It is not our intention to demand that system designers prioritize intimate privacy threats ahead of all other design considerations. Rather, by bringing to the fore considerations about an underspecified privacy threat, we suggest that designers take into account the concerns described in this article during systems design, understanding that they will need to be weighed against other goals and requirements.

Figure 1 offers a heuristic for understanding common relationships between the features of intimate threats and their design implications. It summarizes the four common features of intimate threats we have described in the previous section, and how recognition of these features might inform more thoughtful design. We offer the heuristic not as a definitive, exhaustive list, but as an analytic guide for assessing the risks of intimate threats, the resources they bring to bear, and potential remediations against them. We also do not claim that any one feature is necessarily *exclusive* to intimate threats—indeed, some are shared by other contexts of insider attack, for example—but we believe the constellation of features we describe here is distinctive enough to merit treating intimate threats as their own class of privacy threat.

With all this in mind, we offer the following general design considerations, drawing from the common features we have enumerated, for system designers to help prevent and ameliorate intimate threats.

### Implication 1: Recognize privacy in intimate contexts as a balance among multiple interests and values

As we have discussed, some degree of monitoring is inevitable, desirable, and perhaps even necessary in intimate relationships. Designing for intimate privacy means acknowledging this and finding ways to balance among legitimate interests in privacy protection, safety and caretaking, trust and closeness, and authority—while also acknowledging that these interests may carry very different weights across different relationships, cultures, and points in time.

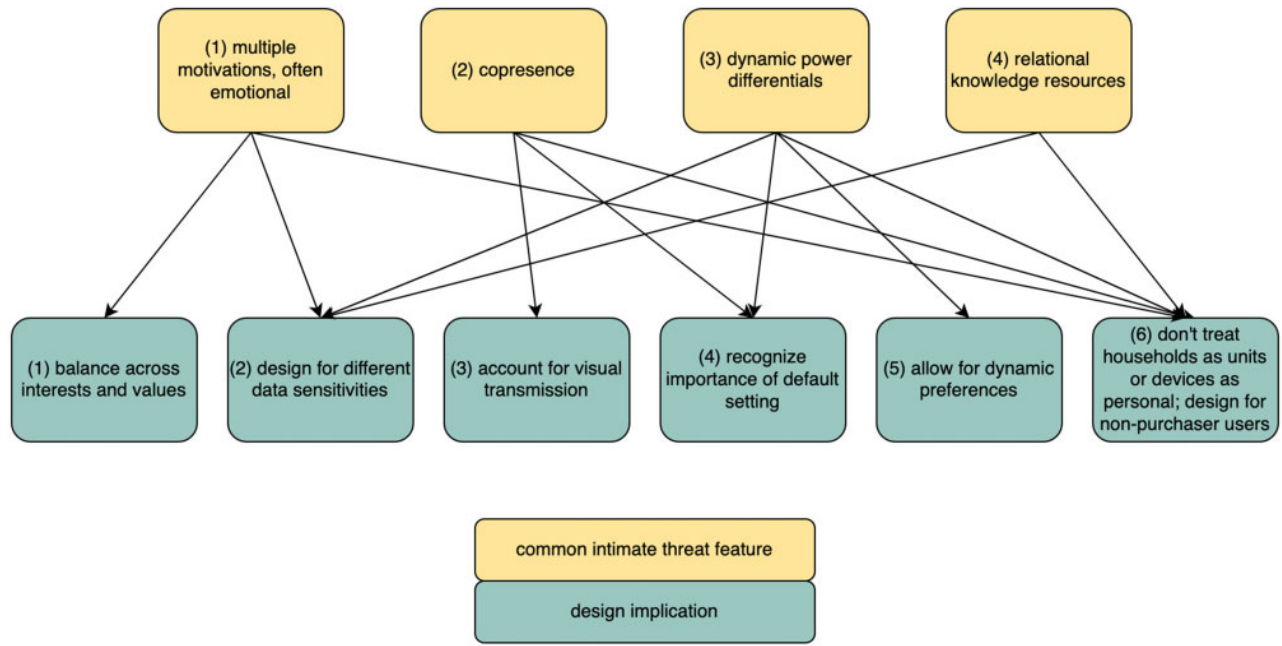

**Figure 1**: Some common features of intimate threats and their design implications. The list offered here is non-exhaustive but offers a heuristic for thinking about designing with intimate threats in mind. The arrows in the diagram are intended to indicate what design considerations we consider to be especially salient in the presence of particular threat features.

There are some good examples of tools that strike this difficult balance well. For example, License+ is a teen driver monitoring app that aims to provide parents with "just the right amount of information so they can stay up-to-date…and the driver doesn't feel spied on" [107]. This is accomplished by giving parents access to a teen's city-level location data (not finer-grained GPS coordinates) and bounding use of the app to 100 total hours—enough to coach new drivers into good practices, but not enough to surveil them indefinitely. The design of the app recognizes parents' legitimate interests in their children's safety, but balances that against a teen's desire for privacy. Balancing these competing interests is difficult and context-specific; a good first step is simply to specify and acknowledge the values at stake and how they may be in tension with one another.

### Implication 2: Recognize different data sensitivities to intimate threats

Intimate threats may have impacts on the types of data that require extra protection. For example, location data, friends lists, calendar data, and communications are likely targets for an intimate attacker, who wants to know where the victim is and with whom they are talking [108]. Data that are normally considered sensitive (like financial account numbers and identification information) may or may not be as salient given an intimate attacker's motivations.[4]

Intimate attacks can also intersect with more conventional privacy concerns in non-obvious ways. For example, a victim might regularly receive confidential information in the course of their work: for example, as a doctor, lawyer, or therapist. This information might be accessed as part of an intimate attack, and then either disclosed or used as a coercive lever. Traditional threat models often fail in these contexts, and system designers should consider whether they have addressed threats against sensitive data from an intimate perspective, and not just a financial or political one.

### Implication 3: Evaluate what information may be inadvertently transmitted through visual display

As we have discussed, some intimate privacy threats occur by virtue of copresence between victim, attacker, and device. Designers should be attentive to what information is displayed visually on the user interface, recognizing that this can be a vector for a privacy breach. Such disclosures are likely to be inadvertent on the part of the user, and information may be actively or passively received by an intimate adversary. In either case, these common disclosures demonstrate how a device can inadvertently divulge information that its owner may prefer to keep private [110, 111].

For example, most mobile operating systems display the content and sender of text messages on the lock screen of a device by default, as well as playing an audio alert indicating that a message has been received. When another app is in use, iOS displays an incoming text as a notification at the top of the screen. Operating systems on laptop and desktop computers also commonly display headers of incoming emails, text messages, Twitter direct messages, or other forms of contacts on-screen as they come in. Such design choices, while intended to be convenient for users, often lead to disclosures of private information when a device screen is in view of another person. Because such notifications typically "push" instantly upon receipt of a message, they further reduce a user's capacity to manage her privacy temporally (e.g., receiving notifications when she can view them without the presence of an intimate).

Targeted online ads are another example. Much can be inferred about a person's interests and characteristics based on what ads are targeted to them. If a browser is shared between family members, ads may "follow" a user around the web and could, quite by accident, reveal to subsequent users what sorts of things a parent, child, or intimate partner has been searching for [112]. Predictive features can also be revealing if viewed visually. Many systems predict recipients of communications based on previous activity. iOS suggests recipients of texts based, presumably, on the frequency and recency of contacts with them; Google Inbox's interface similarly suggests frequent contacts when new messages are being composed. Predictive text within conversations can be similarly disclosive. For instance, iOS's personalized autocorrect dictionary learns and suggests proper names, such as contacts' names, which may reveal information about a user's communication *patterns* should another person see or use the device.

---

4　However, see findings suggesting that users have similar reported data-type sensitivities for insider (i.e., friend) access as stranger access [109].

When Firefox first introduced its private browsing feature, it was indicated by a purple bar across the browser window. This could by easily noticed from across the room, making it harder for someone in the same physical location as another to use the feature without it being obvious. Firefox has since changed the indication to a more discreet purple circle in the upper-right corner of the browser window. Apple's Safari is still problematic: when the user enables private browsing, the normally white address bar turns grey. A better design would be to allow the user to disable any visual indication of private browsing. Other researchers propose inconspicuous forms of data entry, including haptic modalities and coded information [113].

There are other contexts in which designers are attentive to visual privacy invasion without significantly impeding usability. ATMs are designed with keyboard blockers to allow PINs to be entered privately; most websites mask entered passwords as bullets to prevent them from being revealed to screen onlookers [97]. Security mechanisms for the visually impaired are particularly attentive to visual and aural eavesdropping; Azenkot et al. [114] developed a multi-touch authentication method to protect against these risks. Google researchers developed a facial–recognition–based security feature to alert smartphone users when a gaze other than the user's is detected looking at the screen [115]. Mac laptops turn off visual notifications when the display is being projected externally, in recognition of the fact that users showing their screen to a group likely do not want their private messages displayed. The NCAA built a "boss button" into its March Madness streaming site: if employees are watching basketball at work and the boss walks by, they can click the button and an unremarkable spreadsheet pops up temporarily to create the appearance of productivity [116]. Similar "escape" features appear on some intimate partner violence resource sites, to take the user to a generic webpage should an abuser walk into the room.

## Implication 4: Recognize the importance of default-setting and the "blank slate" problem

Privacy defaults are important in all contexts: in general, people are unlikely to change the default settings of a system or service, due to inattention, lack of awareness, or technical difficulty. But in intimate contexts, default-setting is even more important. The launch of Google Buzz in 2010 serves as an illustrative example of the power of defaults. This early microblogging service automatically created a circle of friends for new users based on their most frequent email and chat contacts in Gmail. This was a privacy disaster for many in (or having left) abusive relationships, in some cases leading to physical endangerment for abuse survivors [117]. Having a different default would have prevented this problem from arising.

Furthermore: in intimate contexts, even when disclosive settings can be manually overridden by the user, overriding a default can *itself* create suspicion that the user has something to hide [6]. In most contexts, if an attacker compromises an account or device, we advise the victim to change the access credentials, to open a new account, to cut up the credit card, or otherwise to insulate themselves from the invasion. But in intimate contexts, this is fraught advice, given its limited effectiveness and the risks of escalation. Changing settings to protect one's privacy might be a dangerous "tell," signaling that the victim does not trust the attacker. Therefore, even taking steps to protect oneself against privacy invasion can create danger.

We call this the "blank slate" problem: removing an attacker's access to data, without plausible deniability, may be the worst thing one can do. In abusive relationships, enabling additional privacy protections may result in escalating levels of abuse, thus further endangering the victim. The assumption that one has nothing to hide and thus will not take steps to protect their privacy is an example of what Marques et al. [19] term "performative vulnerability": taking too many affirmative steps to prevent another's access suggests a lack of trust. The same can be true of explicit conversations about access expectations. Stuart Schechter points out that "least privilege may be among the most sacred and respected principles of information security, but starting a conversation on appropriate use of household resources by informing children that their privileges are restricted to a prescribed set of allowable behaviors is a sure way to incite or escalate a conflict" [118]. More generally, the lack of trust that is often the foundation of an effective privacy policy can actively erode relations between intimate partners, family, and friends.

In this vein, Griggio et al. [55] advocate for allowing "discreet changes to privacy preferences" to avoid the unwanted communicative aspect of turning on a privacy setting against an intimate partner. Apple's iOS offers an example in clear contravention of this advice. When Alice takes an affirmative step to stop sharing her location information with Bob, Bob is explicitly notified in the iMessage chat that "Alice has stopped sharing location with you." This setting, which is not to our knowledge overridable by users, may pose real danger to users trying to protect themselves from intimate threats.

## Implication 5: Recognize that privacy and sharing preferences are dynamic

System designers should take into account that sharing preferences will change: couples will break up, children will grow up, roommates will move in and out [12, 100]. Over the course of relationships, intimates' uses of technology and their sharing and privacy preferences are likely to evolve to best suit their current relational aims. And more broadly, sharing norms and societal privacy expectations change over time. Technologies that fail to allow for change run the risk of ossifying outdated privacy expectations to the detriment of users' current preferences.

This fluidity has two primary implications for designers. First, to the greatest extent possible, systems should accommodate changes to preferences. The ability to make discreet changes to privacy settings, discussed above in implication 4, is one aspect of this flexibility; designers may also take steps to avoid the ossification of sharing preferences, for example, by periodically prompting users to ensure that preferences have not changed and that they are aware of what is being shared.

Second, intimate privacy threats often become most salient at discrete moments of relationship transition [7, 54]. Systems should support users when they try to separate joint accounts and help account owners monitor their accounts for login attempts by ex-partners. This means recognizing and accounting for changing privacy preferences over time, not just at the discrete moment of account setup. Facebook has taken some positive steps in this regard. When a user changes their relationship status on the site to indicate a breakup, Facebook proactively displays a prompt asking them if they wish to adjust privacy settings with respect to the ex-partner (for example, hiding future posts from their ex, untagging their ex in past posts) [119].

### Implication 6: Realize that households are not units; devices are not personal; the purchaser of a product is not its only user

System designers build in assumptions about intrafamilial privacy expectations, and often treat a household as a "unit" for purposes of information sharing. These assumptions are incorrect if a privacy threat comes from within one's own household. Often when an account is shared (a cell phone family plan, a TV streaming subscription, a smart home service, health insurance coverage), all users' data associated with that account is accessible to whoever is responsible for payment. But this need not be the case. For example, a single Netflix account is regularly shared amongst an entire household, even though individual users may watch content on different screens. Netflix's security architecture supports multiple profiles in one account, but there is no privacy between them [120]. On the other hand, YouTube TV also supports multiple profiles, but allows those profiles to be individually password-protected, enabling people in a household to better balance their individual needs for sharing and privacy (see also [121]).

Similar failures may occur when households share common channels for information transmission. This often arises when information collected from Internet use is transferred to the real world. Unsolicited email is delivered to an individual email box, while unsolicited paper is delivered to a (shared) household physical mailbox. This difference was illustrated in a widely read privacy anecdote where Target Corporation deduced that a young woman was pregnant and sent her a paper flyer with baby-related offers, alerting the woman's father to the pregnancy before she told him [122]. Similarly, Pakistani law enforcement assures legal adult victims of cyber-harassment of confidentiality when they register complaints online, but then delivers further communication to the victim's house—which is predominantly a family home [123]. A similar issue can occur in cars, which increasingly offer a Bluetooth interface to connect with the driver's phone—and may announce when and from whom a driver receives a text message or a phone call, despite the fact that the car is often a shared space. Smart home technologies present particular challenges in this regard; taking steps like providing visible indicators of data capture (e.g., lights that flash when audio or video is being recorded) can be one way to allow multiple users with divergent privacy preferences to better protect their privacy interests vis-à-vis one another [100, 124].

The converse of the above assumption is that devices considered "personal" are used by only one person. But abundant research demonstrates that this is often not the case, and that device sharing can facilitate unwanted information disclosure [16, 125]. For example, many user interfaces offer seamless integration of content across devices, under the apparent assumption that each of a user's devices will be used by that user alone. For instance, if a user has an iCloud account to which two devices—say, an iPhone and an iPad—are registered, iOS will by default sync iMessages across both devices. But in a family, devices are often shared, rather than being used solely by one iCloud registrant. The seamlessness of this integration fails to realistically reflect typical device usage patterns, and can facilitate inadvertent disclosures in so doing.

System designers should design with *all* potential users' privacy in mind. Companies have a market incentive to build devices for the benefit of the paying customer. But if the use of a device increases privacy risk to another person who is *not* the direct customer, the interests of that person must be protected as well.

Most fundamentally, data access should not be covert. An app to monitor a loved one's cell phone that has no visible icon seems more likely to be used without consent than one that reveals itself [48]. Another approach to preventing covert access is to leave an "access trail" letting users know when their data has been viewed. For example, in Norway, all salary data is public—but searches can't be conducted anonymously, and people can see who has viewed their salary [126]. Facebook employees similarly get a "Sauron alert" from the company if a colleague accesses their account [127]. Though measures like these do not prevent access, they do prevent *covert* access, making it more likely that privacy preferences will be governed by social and relational norms. Improving the discoverability of monitoring is not a silver bullet to the problem of intimate privacy threat, but it can be a useful tool to help prevent and provide recourse against unwanted surveillance.

## Conclusion

Data gathering in intimate relationships is likely to increase in the near future, both due to the increased digital traces on social media and the proliferation of data-gathering devices in homes. An enormous number of consumer IoT products are explicitly marketed for the protection, supervision, and care of intimates [28]. Even IoT devices that are not specifically so marketed often allow us to draw inferences about an intimate's activities, and often without their awareness [128]: web-enabled security cameras that capture the behaviors of anyone in the home [129], or the sleep tracker that records the activities of anyone using the bed [52]. The growth of this consumer market and the continuing normalization of monitoring across intimate relationships makes this a class of threats to be taken seriously.

There are some signs that intimate threats are beginning to be recognized by the tech industry. For example, Kaspersky recently announced an effort to alert users to the presence of stalkerware apps covertly installed on Android products [130], and Google made some efforts to scrub similar apps from its Play Store following research about their prevalence [48]. We take heart at these developments, but suggest that consideration of intimate threat models should be more thoroughly integrated into system design broadly, rather than only in response to the most egregious apps for covert intimate monitoring.

Addressing these threats not only extends the field of cybersecurity to meet the needs of vulnerable communities, but also brings it into fruitful dialogue with other disciplines and modes of inquiry. It requires an integrated sociotechnical approach to understanding privacy. It requires focusing our attention both on new problems and new tools for addressing them, taking seriously the social and cultural sites within which technologies and users are situated, and acknowledging the full range of harms privacy threats can pose. It requires thinking more broadly about how we design secure systems.

By recognizing the class of intimate threats and characterizing their common features, we can begin to articulate design principles to address them.

## Acknowledgments

This work was supported by the National Science Foundation [CNS-1916096]. The authors would like to thank Kendra Albert, Nighat Dad, Jessica Dawson, Shauna Dillavou, Beth Friedman, Vicki Laidler, Damon McCoy, Barath Raghavan, Stuart Schechter, and Adam Shostack for their helpful comments on a draft version of this article, as well as Clara Berridge, Rahul Chatterjee, Nicki Dell, Periwinkle Doerfler, Diana Freed, Jodi Halpern, Sam Havron, Lauren Kilgour, Damon McCoy, Tom Ristenpart, and Luke Stark for prior research collaborations that informed this work. Finally, the authors would like to thank the interdisciplinary workshop on Security and Human Behavior (SHB) for sparking this collaboration.